\documentclass[11pt]{article}
\usepackage{amssymb}
\usepackage{epsfig}
\newlength{\bredde}
\def\slash#1{\settowidth{\bredde}{$#1$}\ifmmode\,\raisebox{.15ex}{/}
\hspace*{-\bredde} #1\else$\,\raisebox{.15ex}{/}\hspace*{-\bredde} #1$\fi}
\newcommand{\be}{\begin{equation}}
\newcommand{\ee}{\end{equation}}
\newcommand{\bea}{\begin{eqnarray}}
\newcommand{\eea}{\end{eqnarray}}
\newcommand{\nn}{\nonumber}
\newcommand{\al}{\alpha}
\newcommand{\la}{\lambda}

\newcommand{\ep}{\epsilon}

\newcommand{\sect}[1]{\setcounter{equation}{0}\section{#1}}

\def\im{{\Im\mbox{m}}}

\begin{document}
\topmargin -1.4cm
\oddsidemargin -0.8cm
\evensidemargin -0.8cm
\title{\Large{{\bf
Universal random matrix correlations of ratios of\\
characteristic polynomials at the spectral edges
}}}

\vspace{1.5cm}
\author{~\\{\sc G. Akemann}$^1$ and {\sc Y.V. Fyodorov}$^{2}$\\~\\
$^1$Service de Physique Th\'eorique, CEA/DSM/SPh$\!$T Saclay\\
Unit\'e de recherche associ\'ee au CNRS\\
F-91191 Gif-sur-Yvette Cedex, France\\~\\
$^2$Department of Mathematical Sciences, Brunel University\\
Uxbridge, UB8 3PH, United Kingdom}

\date{}
\maketitle
\vfill
\begin{abstract}
It has been shown recently \cite{FS} that Cauchy transforms of
orthogonal polynomials appear naturally in general correlation
functions containing ratios of characteristic polynomials of
random $N\times N$ Hermitian matrices.
Our main goal is to investigate
the issue of universality of large N asymptotics for those
Cauchy transforms for a wide class of weight functions.
Our analysis covers three different scaling regimes:
the "hard edge", the "bulk"
and the "soft edge" of the spectrum, thus extending the earlier
results known for the bulk. The principal tool
is to show that for finite matrix size $N$
the auxiliary "wave functions" associated with the Cauchy transforms
obey the same second order differential equation
as those associated with the orthogonal polynomials themselves.
\end{abstract}
\vfill

\begin{flushleft}
SPhT T03/036\\
\end{flushleft}
\thispagestyle{empty}
\newpage

\renewcommand{\thefootnote}{\arabic{footnote}}
\setcounter{footnote}{0}

\sect{Introduction}\label{intro}

Characteristic polynomials are important objects in Random Matrix Theory
(RMT) and enjoy a wide range of applications in various fields of
physics and mathematics. To name only a few, we mention the
applications to Quantum Chromodynamics (QCD) \cite{Jac}, Quantum Chaos
\cite{AS} as well as their role in elucidating properties
 of the Riemann zeta function along the critical line \cite{KS,BH}.
In fact, products of ratios of the characteristic polynomials
at different values of the spectral parameter
provide the most general generating functional for all spectral
correlation functions in RMT. Characteristic polynomials
are also useful in providing a direct link between the RMT description
and the effective field theory such as Chiral Perturbation Theory
($\chi$PT) in the application to QCD \cite{AD,DOTV,KimJac}.
There, products of the ratios of characteristic polynomials enjoy a
direct interpretation as partitions functions
including fermionic and bosonic degrees of freedom.
However, these objects are very difficult to obtain in
full generality from the finite volume limit of $\chi$PT, where one has to
parameterize and calculate supergroup integrals \cite{DOTV}.
Alternatively, one has to perform replica calculations exactly
\cite{KimJac}, a possibility opened due to recent progress
in understanding the analytical structure of the moments \cite{EK}.
From this side, correlation functions of
characteristic polynomials recently calculated
within  RMT \cite{FSch,FS} provide the result yet to be obtained
\cite{FA} directly from the field-theoretic calculations.

An important issue for all RMT calculations is the question of
universality. By this one usually understands
that the predictions of RMT for a given object are insensitive to the
details of the particular models.
When this property holds, it must be a consequence of the common
underlying symmetry of that class of the models. This
type of argument validates applying RMT results
to systems of varying nature in all the diverse situations
mentioned above.

The character of universality depends essentially
on the scale in the spectrum one is interested in, as well as
on the nature of the singularity at the support of the spectrum
(the so-called "edges" of the spectrum).
In the following we will be only concerned with the
scales comparable with the mean level spacing.
This issue is widely known as that of "microscopic" universality.
Suppose first one is interested in such quantities as
the correlation function of eigenvalue densities, or the distribution
of the largest/smallest eigenvalue, or some related objects.
Then restricting oneself to the class of
random matrix measures invariant with respect to the change of basis
(here the unitary invariant ensembles)
the quantity of interest can be expressed as a
determinant containing kernels made of polynomials
orthogonal with respect to that measure \cite{Mehta}.
In this way studying universality amounts mainly to
investigating asymptotics of the orthogonal polynomials in various
regimes, and
is well established by now both on physical
\cite{BZ,ADMN,KF,DN,KF97b,ADMNII,AVII}
as well as mathematical \cite{PS,Deift1,BI,Deift2} level of rigor.

On the other hand, more general correlation functions containing
products and ratios of the characteristic polynomials
(in particular, those used to generate all
the eigenvalue correlations by differentiation and taking the
discontinuity of the resolvents) attracted considerable
attention only recently \cite{KimJac,FSch,FS,FA,FSuniv}.
While for the products only the universality follows quite straightforwardly
from the mentioned results \cite{BZ,ADMN,KF,DN,PS,Deift1,BI,Deift2}
(see also \cite{Poul}), the universality of expressions
containing inverse powers or ratios was properly understood only in the bulk
of the spectrum \cite{FSuniv}.

There are several possibilities to calculate
correlation functions containing ratios of characteristic
polynomials. For the case of the Gaussian measure
the only tool until quite recently was the exploitation of anticommuting
Grassmann variables (the method of "supersymmetry" due to Efetov \cite{Efetov},
or its variants \cite{Guhr,my,AS,FSch,FS}).
An alternative for general unitary-invariant measures
was discovered quite recently by Fyodorov and Strahov
\cite{FS,FSuniv} and resorts again to the orthogonal polynomial method.
An important new feature, however, is that as soon as inverse powers of
characteristic polynomials are taken into consideration,
 one has to deal not only with the orthogonal polynomials
but also with their so-called Cauchy transforms. More precisely,
a very general result - valid for all invariant ensembles of
finite-size complex Hermitian matrices \cite{FS} - reduces arbitrary
correlation function containing
ratios of characteristic polynomials to a determinant made of
orthogonal polynomials and their Cauchy transforms.
The universality of the corresponding expression in the
microscopic limit in the "bulk" of the spectrum was rigorously
demonstrated by Strahov and Fyodorov \cite{FSuniv} by using
the Riemann-Hilbert approach.
Those authors however considered neither the "soft" nor the "hard" edge of the
spectrum, the latter being a characteristic feature of the so-called "chiral"
version of the unitary ensemble \cite{chir,Jac}.

Our goal in the present paper is to employ an alternative
method and to obtain universal asymptotics
for all three regimes, including spectral edges of both types.
This is particularly important
in view of the application to QCD and the role of
characteristic polynomials played there (see. e.g. in \cite{KimJac,FA}
and references therein).
The method we are going to exploit to extract universal features
of the correlation functions
is based on the works by Kanzieper and Freilikher \cite{KFrev}.
The clear advantage of this technique is its explicit form and
conceptual simplicity,
although it lacks the rigor and vigor of the Riemann-Hilbert
technique (and in fact makes use of some facts proved by the latter).
First we will show the existence
of an exact second order differential equation satisfied by
the Cauchy transform of the orthogonal polynomials
for a large class of measures. Then by
considering the microscopic large-$N$ limit
of that differential equation we can demonstrate their universality
at the origin ("hard edge"), in the bulk and at the soft
edges of the spectrum.
The procedure follows exactly
the same lines as for the orthogonal polynomials themselves (see
\cite{KFrev}), since in both cases we have to
deal with the very same equation.
The universality of asymptotics for the general expressions
\cite{FS} of arbitrary ratios of characteristic polynomials
then follow. It is necessary to mention that the
method implicitly uses an assumption of the existence of a
single-valued function arising in a certain limit
for recursion coefficients relating orthogonal polynomials of
different order. This was indeed shown to be true
in the case of a spectrum supported on a single interval
in the framework of the Riemann-Hilbert
technique.

The article is organized as follows. In sect.
\ref{defs} we give the main definitions and recall the main
result of \cite{FS}.
In the following sect. \ref{diff} we derive a differential equation
for the wave functions associated with the Cauchy transforms for
finite $N$
and show that it is identical to that for the wave functions of
orthogonal polynomials.
For convenience we treat the unitary ensemble (UE) first and
translate to the  chiral unitary ensemble (chUE) later on.
Sect. \ref{univ} deals with the large-$N$ limit
yielding universal results for both
unitary and chiral unitary ensemble in all three microscopic regimes.
We close with the discussion and open questions in sect. \ref{con}.

\sect{Definitions}\label{defs}

In this section we introduce the main objects we wish to study in the
present paper.
The matrix model partition function with respect to which we calculate
expectation values is given by the following integral over the
(unitary invariant) Haar measure
for matrices $H$:
\be
{\cal Z}_N \ \sim\ \int_D [dH]_{N\times N}\  w(H)
     \ \sim\ \int_D \prod_{k=1}^N d\la_k\  w(\la_k) \Delta_N(\Lambda)^2 \ .
\label{defZN}
\ee
The matrix $H$ can be either complex Hermitian, representing
a general unitary invariant ensemble without chiral symmetry,
or it can be decomposed into a product of complex rectangular
matrices $W$ of size $N\times (N+\nu)$, $H=W^\dagger W$,
representing the chUE. In the first case the matrices $H$
can be diagonalized by a unitary transformation with
real eigenvalues
$\Lambda=\mbox{diag}(\la_1,\ldots,\la_N)$ are distributed according to a
weight function $w(\la)$ with support on ${\mathbb R}$.
In the second case we deal rather with real positive singular values
of the matrices $W$ with the support being
${\mathbb R}_+$, respectively. We have also employed the
notation $\Delta_N(\Lambda)\equiv\prod_{i>j}^N(\la_i-\la_j)$
for the Vandermonde determinant.
While the results cited in this section hold
for general weight functions with convergent moments on some domain
$D\subseteq \mathbb{R}$,
we will concentrate mainly on weights of the form $w(\la)=\exp[-NV(\la)]$
in the following sections. The potential $V(x)$ may contain
both an (even) polynomial part and a logarithmic part.

We will be interested in extracting the large-$N$ universal behavior
of the correlation functions defined via expectation values of
arbitrary ratios of characteristic polynomials:
\be
{\cal K}_N(M_f,M_b)\ \equiv\
\left\langle
\frac{\prod\limits_{j=1}^{L}\det\left(m_j^{f}{\bf 1}_N+H\right)}
{\prod\limits_{j=1}^{M}\det\left(m_j^{b}{\bf 1}_N+H\right)}
\right\rangle .
\label{defKN}
\ee

In the context of QCD the matrices
$M_f=\mbox{diag}\left(m^{f}_1,...,m^{f}_L\right)\,\,,\,\,
M_b=\mbox{diag}\left(m^{b}_1,...,m^{b}_M\right)$ are said to contain the
fermionic $(f)$ and bosonic $(b)$ mass terms. The expectation value of
ratios of characteristic polynomials is known as the supersymmetric partition
function. It may be used to calculate correlation functions
of the eigenvalue densities and serves as a general generating function for
multi-point resolvent operators.

In order to give rise to positive definite partition functions in the case
of the UE the masses have to be chosen purely imaginary and occur in pairs of
complex conjugates. In this way only positive definite terms
appear, $\det(im+H)\det(-im+H)=\prod_k^N(m^2+\la_k^2)$. In the chUE each
determinant factor is positive provided we parameterize
$\det(m^2+W^\dagger W)=\prod_k^N(m^2+\la_k),\ \la_k\geq0$.
At the same time, from the mathematical point of view
there is no need to impose any restrictions on masses
in eq. (\ref{defKN}),
apart from requiring infinitesimal imaginary parts for all $m_j^b$.

As it has been shown in \cite{FS} for general weight functions and
finite size $N$ the generating function eq. (\ref{defKN})
can be written as a determinant of
size $M+L$ containing polynomials orthogonal with respect to the weight
$w(\la)$. Let us define the monic orthogonal polynomials
$\pi_n(\la)=\la^n+{\cal O}(\la^{n-1})$
\be
\int_{D}d\la\, w(\la)
\pi_n(\la)\pi_m(\la)=\delta_{nm}f_n
\label{defOP}
\ee
with normalization factors $f_n$ uniquely defined through the Gram-Schmidt
orthogonalization.\footnote{ For weight functions on $D=\mathbb{R}$
the corresponding norms can be calculated recursively solving
$n f_{n-1}=-\int_{-\infty}^{\infty}\! d\la~ w'(\la) \pi_n(\la)\pi_{n-1}(\la)$
for even potentials plus an additional equation for potentials without parity.}
The second building block are the Cauchy transforms of those polynomials
which appear only as long as $M>0$,
\be
\vartheta_n(\epsilon)\ \equiv\ \frac{1}{2\pi i}\int_{D}
d\la\ \frac{w(\la)}{\la-\epsilon}\pi_n(\la)\ ,\ \ \Im m(\epsilon)\neq 0 \ .
\label{Cauchy}
\ee
The result derived in \cite{FS} then
reads \footnote{The object eq. (\ref{KN}) can be given a meaning for real
values of the mass parameters $m_l^b$ as well, by choosing the principle value
in eq. (\ref{Cauchy}). However, to use it as a generating functional we are
precisely interested in its discontinuities in the complex plane.}
\be
 {\cal K}(M_f,M_b)\propto \frac{1}{\Delta(M_b)\triangle(\hat{M}_f)}
\,\mbox{det}\left|
\begin{array}{cccc} \vartheta_{N-M}\left(-m^{b}_1\right) &
\vartheta_{N-M+1}\left(-m^{b}_1\right) &
\ldots & \vartheta_{N+L-1}\left(-m^{b}_1\right) \\
\vdots &\vdots &  &\vdots \\ \vartheta_{N-M}\left(-m^{b}_M\right) &
\vartheta_{N-M+1}\left(-m^{b}_M\right) &
\ldots & \vartheta_{N+L-1}\left(-m^{b}_M\right) \\
\pi_{N-M}\left(-m^{f}_1\right) &
\pi_{N-M+1}\left(-m^{f}_1\right) &
 \ldots & \pi_{N+L-1}\left(-m^{f}_1\right) \\
\vdots &\vdots &  &\vdots \\  \pi_{N-M}\left(-m^{f}_L\right) &
\pi_{N-M+1}\left(-m^{f}_L\right) & \ldots &
\pi_{N+L-1}\left(-m^{f}_L\right)
\end{array}\right|.
\label{KN}
\ee

\sect{A differential equation for the orthogonal polynomials and their
Cauchy transforms}\label{diff}

As is well known \cite{BZ,KFrev,BI} it is convenient to
relate to the system of orthogonal polynomials $\pi_n(\lambda)$
a system of associated objects:
\be
\psi_n(\la)\ \equiv\ \exp\left[-\frac{N}{2}V_\al(\la)\right]\
\frac{\pi_n(\la)}{\sqrt{f_n}}\ .
\label{psi}
\ee
frequently called the "wave functions".
The name has its origin in two properties: (i) the functions
$\psi_n(\lambda)$ obey a closed second order
differential equation for finite $N$ which
can be interpreted as a kind of one-dimensional Schroedinger equation
and (ii) the functions are orthonormal. The first property can be
shown, in particular, by the formalism introduced by Kanzieper and Freilikher
\cite{KFrev}, to which we refer an interested reader for a detailed
review. Below we exploit the same formalism to infer properties
of the Cauchy transforms of the orthonormal polynomials, after
suitable redefinition of the wavefunctions.
This will enable us to treat the Cauchy transforms
on equal footing with the polynomials themselves when extracting the
universal properties in the microscopic large-$N$ limit.
The wavefunctions of the orthogonal polynomials
and of their Cauchy transforms will be then two independent solutions of the
second order differential equation, with the latter being singular at the
origin.

Due to an intimate relation between the polynomials of the
UE and the chUE for general weights $w(\la)$
we can restrict ourselves to treat only
the UE in greater detail in the following subsection.
Once our results are established those for the chUE
will follow rather straightforwardly, as discussed in
in the subsection \ref{chUE}.

\subsection{The Unitary Ensemble}\label{UE}

For the UE the eigenvalues of the complex Hermitian matrix $H$ are
distributed along the full real line. To be specific let us introduce the
following parameterization of the weight function
\be
w(\la)\ \equiv\ \la^{2\al}\exp[-NV(\la)] \ \equiv\ \exp[-NV_{\al}(\la)]\ .
\label{weightUE}
\ee
Here we have introduced the potential
\be
V_{\al}(\la) ~\equiv~ V(\la) - \frac{2\al}{N}\ln(\la) ~, \ \
V(\la)=\sum_{k=1}^{d/2} \frac{g_{2k}}{2k} \la^{2k}\ ,
\label{Valpha}
\ee
with an even polynomial part $V(\la)$ of arbitrary but fixed even degree $d$.
In order to apply the formalism of \cite{KFrev} we switch from monic
polynomials $\pi_n(x)$ to orthonormal polynomials,
\be
\delta_{kl} ~=~
\int_{-\infty}^{\infty}\! d\la~ w(\la) p_k(\la)
      p_l(\la) ~, \ \ p_k(\la)= \pi_k(\la)/\sqrt{f_k}\ .
\label{OP}
\ee
 The potential in eq. (\ref{Valpha}) being even, the polynomials
possess the following parity property:
\be
p_n(-\la)\ =\ (-1)^n \ p_n(\la) \ .
\label{parity}
\ee
As is well known (see e.g. \cite{BI,KFrev} and references therein),
all orthogonal polynomials on the real line or a subset of it
satisfy a three-step recursion relation
\be
\la p_{n}(\la) ~=~ c_{n+1}p_{n+1}(\la) + c_{n}p_{n-1}(\la) ~,
\label{rec}
\ee
where the middle term vanishes because of the parity.
The recursion coefficients can be obtained from the normalization
factors $f_n$ of the polynomials as:
\be
c_n\ =\ \sqrt{\frac{f_n}{f_{n-1}}} \ .
\label{reccoeff}
\ee
The second operation, which together with  the multiplication
by $\lambda$ in eq. (\ref{rec}) closes on the ring of orthogonal
polynomials, is differentiation over $\lambda$.
Following  \cite{KFrev} it can be most conveniently written as
\be
p_n'(\la) ~\equiv~ {A}_n(\la)p_{n-1}(\la) ~-~
                       {B}_n(\la)p_{n}(\la) ~.
\label{Pprime}
\ee
Here, the functions $A_n(\la)$ and $B_n(\la)$ are defined as
\bea
{A}_n(\la)&\equiv&N~ c_n\int_{-\infty}^{\infty}\! d\eta~
\frac{w(\eta)}{\eta-\la}
     [V_{\al}'(\eta) - V_{\al}'(\la)]\ p_n(\eta)^2  \nn\\
&=&N~ c_n\int_{-\infty}^{\infty}\! d\eta~
\frac{w(\eta)}{\eta-\la}
     [V'(\eta) - V'(\la)]\ p_n(\eta)^2 ~,\label{An}\\
{B}_n(\la)&\equiv&N~ c_n\int_{-\infty}^{\infty}\! d\eta~
\frac{w(\eta)}{\eta-\la}
  [V_{\al}'(\eta) - V_{\al}'(\la)]\ p_n(\eta)p_{n-1}(\eta) \nonumber\\
         &\equiv& {B}_n^{\mbox{\scriptsize reg}}(\la)
            ~+~ (1-(-1)^n)\frac{\al}{\la} ~.
\label{Bn}
\eea
Because of parity the singular
$\al$-dependent term drops out in  $A_n(\la)$. Inserting
the explicit form of the potential from eq. (\ref{Valpha})
it can be easily seen that
$A_n(\la)$ and the regular part $B_n^{\mbox{\scriptsize reg}}(\la)$
of $B_n(\la)$ are polynomials of degree $d-2$ and $d-3$ respectively.
From their definition one can easily show
the following identity to be valid \cite{KF}:
\be
{B}_n(\la) + {B}_{n-1}(\la) ~+~  N~ V_{\al}'(\la)
~=~ \frac{\la}{c_{n-1}}{A}_{n-1}(\la)~.
\label{ABV}
\ee
This identity, when combined with the two properties of the
polynomials - the
recursion relation eq. (\ref{rec}) and the differentiation eq.
(\ref{Pprime}) - suffices to show that the following second order differential
equation is satisfied by the wavefunctions defined in Eq.(\ref{psi}):.
\be
\psi_n''(\la) - {F}_n(\la)\psi_n'(\la) + {G}_n(\la)
\psi_n(\la) ~=~ 0 ~,
\label{psidiff}
\ee
where
\begin{eqnarray}
{F}_n(\la) &\equiv& \frac{{A}_n'(\la)}{{A}_n(\la)} \label{F}\\
{G}_n(\la) &\equiv& \frac{c_n}{c_{n-1}}{A}_n(\la){A}_{n-1}(\la)
                -\left({B}_n(\la)+\frac{N}{2}V_{\al}'(\la)\right)^2
               \nonumber\\
                &&+\left({B}_n(\la)+\frac{N}{2}V_{\al}'(\la)\right)'
                -\frac{{A}_n'(\la)}{{A}_n(\la)}
                \left({B}_n(\la)+\frac{N}{2}V_{\al}'(\la)\right) \ .
\label{G}
\end{eqnarray}
Here $'$ denotes the derivative with respect to $\lambda$ everywhere.
The differential equation Eq.(\ref{psidiff}) holds for finite-$N$ and
arbitrary weight $w(\la)$ corresponding to a given polynomial
$V(\la)$ of degree $d$ and parameter $\al>-1/2$.
The fact that $\alpha$ can be real will enable us to translate most easily
the results for the UE obtained below to the chUE in the next subsection.

Our goal is to derive a similar differential
equation for the Cauchy transform
at finite-$N$. In terms of the orthonormal polynomials they are given by
\be
h_n(\epsilon) \ \equiv\ \frac{1}{2\pi i}\int_{-\infty}^{+\infty}
d\la\ \frac{w(\la)}{\la-\epsilon}\,p_n(\la)\ ,\ \ \im(\epsilon)\neq 0 \ .
\label{defCauchy}
\ee
This defines in fact two functions, one in the upper and one in the lower half
plane for positive or negative imaginary part, respectively. Each of it
can be extended to the other half plane using the relation
\be
h_n(-\ep) \ =\ (-1)^{n+1}\ h_n(\ep)\ .
\label{hparity}
\ee
which follows easily from the  parity of the polynomials,
eq. (\ref{parity}).
We note that the Cauchy transform has the opposite parity than the
corresponding polynomial.
Let us stress that the transformation (\ref{hparity}) does not relate the
two different functions defined in eq. (\ref{defCauchy}).

In the following we show that the object
\be
\tilde{h}_n(\epsilon)\ \equiv \exp[+NV_\al(\epsilon)]\ h_n(\epsilon)
\label{ht}
\ee
satisfies the same recursion relation eq. (\ref{rec})
and differentiation relation eq. (\ref{Pprime}) as the corresponding
orthogonal polynomial $p_n(\la)$, provided that $n>d-1$.
As an immediate consequence the wavefunction for the Cauchy transform
defined as
\be
\chi_n(\epsilon) \ \equiv\ \exp\left[-\frac{N}{2}V_\al(\ep)\right]
    \, \tilde{h}_n(\ep)
\ =\ \exp\left[+\frac{N}{2}V_\al(\ep)\right]\, h_n(\epsilon)
\ ,
\label{chi}
\ee
satisfies the second order differential equation identical
to Eq. (\ref{psidiff}) at finite-$n$, provided the condition  $n>d-2$.
The corresponding solution of the differential equation
at finite and large-$n$ will then have to be chosen according to the
appropriate boundary condition for $\psi_n$ and $\chi_n$, respectively.

Let us first consider multiplication on $h_n(\ep)$ for $n>0$:
\bea
\ep h_n(\ep) &=& \frac{1}{2\pi i}\int_{-\infty}^{+\infty}
d\la \frac{w(\la)}{\la-\epsilon}\ \ep p_n(\la)\ \ ,\nn\\
&=& \frac{-1}{2\pi i}\int_{-\infty}^{+\infty} d\la\ w(\la) p_n(\la)
\ +\ \frac{1}{2\pi i}\int_{-\infty}^{+\infty}
d\la \frac{w(\la)}{\la-\epsilon}[ c_{n+1}p_{n+1}(\la) + c_{n}p_{n-1}(\la)]\nn\\
&=&  c_{n+1}h_{n+1}(\ep) + c_{n}h_{n-1}(\ep) \ , \ \ n>0\ .
\label{h-rec}
\eea
At the first step we have added and subtracted $\la$ to cancel the denominator
and used the recursion relation eq. (\ref{rec}). At the second step we used
that $p_n(\la)$ is orthogonal to $p_0(\la)=const$ for $n>0$.
The fact that the recursion Eq. (\ref{h-rec}) also holds for
$\tilde{h}_n(\ep)$ as well as for $\chi_n(\epsilon)$ is an
immediate consequence.

The next step is to consider differentiation acting
on $\tilde{h}_n(\ep)$ for $n>d-2$:
\bea
\partial_\ep \tilde{h}_n(\ep) &=&  \frac{1}{2\pi i}\int_{-\infty}^{+\infty}
d\la\ \frac{\mbox{e}^{N(V_\al(\ep)-V_\al(\la))}}{\la-\ep}
\left[ N\left(V_\al'(\ep)-V_\al'(\la)\right)p_n(\la)\ +\ p_n'(\la)\right] \nn\\
&=& \frac{1}{2\pi i}\int_{-\infty}^{+\infty}
    d\la\ \mbox{e}^{N(V_\al(\ep)-V_\al(\la))}
    \left[ N\frac{V'(\ep)-V'(\la)}{\la-\ep}p_n(\la)
           -\frac{2\al}{\ep\la}p_n(\la) \right. \nn\\
&&+\left.\frac{A_n(\la)}{\la-\ep}p_{n-1}(\la)
-\frac{{B}_n^{\mbox{\scriptsize reg}}(\la)
                   +(1-(-1)^n)\al\la^{-1}}{\la-\ep}p_n(\la)\right] \nn\\
&=& \frac{1}{2\pi i}\int_{-\infty}^{+\infty}
    d\la\ \mbox{e}^{N(V_\al(\ep)-V_\al(\la))}
\left[ -\frac{2\al}{\ep\la}p_n(\la)
+\frac{A_n(\ep)}{\la-\ep}p_{n-1}(\la) \right.\nn\\
&&\left. -\ \frac{{B}_n^{\mbox{\scriptsize reg}}(\ep)
                   +(1-(-1)^n)\al\ep^{-1}}{\la-\ep}p_n(\la)
+\frac{\al(1-(-1)^n)}{\ep\la}p_n(\la)\right]\nn\\
&=& A_n(\ep)\ \tilde{h}_{n-1}(\ep)\ -\left[{B}_n^{\mbox{\scriptsize reg}}(\ep)
 +(1-(-1)^n)\frac{\al}{\ep}\right]\ \tilde{h}_{n}(\ep)\ , \ \ n>d-2\ .
\label{hprime}
\eea
Here we first have used the relation $\partial_\ep\frac{1}{\ep-\la}=-
\partial_\la\frac{1}{\ep-\la}$ and performed the integration by parts.
The second step spells out the definition of $V_\al(\la)$ and exploits Eq.
(\ref{Pprime}) for $p_n'(\la)$. We can now use that
$(V'(\ep)-V'(\la))/(\la-\ep)$ is a polynomial of order $d-2$ and thus
orthogonal to $p_n(\la)$ for $n>d-2$. The same argument helps
to rewrite $A_n(\la)/(\la-\ep)= (A_n(\la)-A_n(\ep))/(\la-\ep)+
A_n(\ep)/(\la-\ep)$ and to drop the first term, which is again a polynomial
(this times of the order $d-3$) and thus orthogonal to $p_n(\la)$.
We can repeat the same steps for ${B}_n^{\mbox{\scriptsize reg}}(\la)$
and the poles $1/\la$. The remaining terms $\sim 1/(\la\ep)$
require a little bit more careful treatment.
For $n$ even $p_n(\la)/\la$ is odd and thus the
corresponding integral vanishes identically.
For $n$ odd the last term $\sim (1-(-1)^n)$ gives a
contribution that precisely cancels the first one.
We thus finally arrive at
the same relation for differentiation of
$\tilde{h}_{n}(\ep)$ as that given by eq. (\ref{Pprime}).
Consequently, Eqs. (\ref{h-rec}), (\ref{hprime})
and (\ref{chi}) imply
\be
\chi_n''(\la) - {F}_n(\la)\chi_n'(\la) + {G}_n(\la)
\chi_n(\la) ~=~ 0 ~.
\label{chidiff}
\ee
Thus, the wave functions associated with the
orthogonal polynomials and their Cauchy transforms satisfy
the same differential equation, provided that condition
$n>d-2$ holds.
This constraint is not an obstacle for large but finite-$n$,
for any given potential $V(\la)$ with a fixed finite degree $d$.

The virtue of our analysis at finite $n$ is that it obviously translates
all the universality proofs carried out at the origin, in the bulk and at the
edge of the spectrum for the orthogonal polynomials to their
Cauchy transforms. We will concentrate on the the large-$n$ limit in the next section
\ref{univ}, but first we repeat the above analysis for the chiral
case of chUE.

\subsection{The chiral Unitary Ensemble}\label{chUE}

The eigenvalues for chUE are distributed on the positive real
line ${\mathbb R}_+$ and governed by the following
chiral weight function:
\be
w^{ch}(\la) \ \equiv\ \la^{\al}\exp[-NV^{ch}(\la)]
\ \equiv\ \exp[-NV^{ch}_{\al}(\la)].
\label{chweight}
\ee
 Here the potential $V^{ch}_{\al}(\la)$ is
of degree $d/2$, with $d$ even, and is defined as
\be
V^{ch}_{\al}(\la) \ \equiv\
\sum_{k=1}^{d/2} \frac{g_{2k}}{k} \la^{k}
- \frac{\al}{N}\ln(\la) ~. \ \
\label{Vch}
\ee
The orthogonality relation for the corresponding orthogonal polynomials
denoted as $p^{(\al)}_{k,\ ch}(\la)$ reads
\be
\delta_{kl} ~=~
\int_{0}^{\infty}\! d\la~ w^{ch}(\la) p^{(\al)}_{k,\ ch}(\la)
p^{(\al)}_{l,\ ch}(\la) ~.
\label{OPch}
\ee
Here we explicitly display the dependence on the parameter $\al$.
Substituting $\la^2=y$ one can easily see that
the orthogonal polynomials of the chUE can be obtained from those
of the UE  by simple replacement
of the argument \cite{ADMNII}:
\be
p_{n,\ ch}^{(\al)}(\la^2) \ =\ p_{2n}^{(\al+1/2)}(\la)\ .
\label{OPrel}
\ee
Here we have to appropriately identify the corresponding potential as
\be
V^{ch}(y^2)\ =\ 2\ V(y)\ ,
\label{Vrel}
\ee
which effectively means that the non-chiral polynomials are taken at
values $2N$. In fact, one can look at the relation Eq. (\ref{OPrel})
as a generalization of the well-known relation between
generalized Laguerre and Hermite polynomials.
Consequently, there is no need to repeat the whole analysis of the previous
subsection but one can simple read off the chiral
orthogonal polynomials and their asymptotics
from the same differential equation (\ref{psidiff}) for even $N$ at shifted
value of $\al$.

Let us now show that the Cauchy transforms of the chiral orthogonal
polynomials relate to those of the non-chiral ones as well.
We first define the Cauchy transform for the chiral case as
\be
h_{n,\ ch}^{(\al)}(\ep) \ \equiv\ \frac{1}{2\pi i}\int_{0}^{\infty}
d\la\ \frac{w^{ch}(\la)}{\la-\epsilon}\ p^{(\al)}_{n,\ ch}(\la)\ ,
\ \ \Im m(\epsilon)\neq 0\ .
\label{chCauchy}
\ee
Substituting $\la^2=y$ and
inserting the relation (\ref{OPrel}) we obtain
\bea
h_{n,\ ch}^{(\al)}(\ep) &=& \frac{1}{2\pi i}\int_{-\infty}^{+\infty}
dy\ \frac{y}{y^2-\epsilon}\ \mbox{e}^{-NV^{ch}_{\al}(y^2)}
p^{(\al)}_{n,\ ch}(y^2)
\label{chh}\\
&=& \frac{1}{2\pi i}\int_{-\infty}^{+\infty}
dy\ \frac{1}{y^2-\epsilon}\ \mbox{e}^{-NV^{ch}_{\al+1/2}(y^2)}\
p^{(\al+1/2)}_{2n}(y)\ ,
\label{chh1}
\eea
where we have used the parity of the non-chiral polynomial
$p^{(\al+1/2)}_{2n}(y)$. On the other hand we can rewrite the Cauchy
transform of the non-chiral ensemble eq.
(\ref{Cauchy}) with an even index $2n$ as
\be
h_{2n}^{(\al)}(\ep) \ =\  \frac{1}{2\pi i}\int_{-\infty}^{+\infty}
dy \frac{\ep}{y^2-\ep^2}\ \mbox{e}^{-2NV_{\al}(y)}\
p^{(\al)}_{2n}(y) \ .
\label{chh2}
\ee
Comparing eqs. (\ref{chh1}) and (\ref{chh2}) together with eq. (\ref{Vrel})
we thus arrive at
\be
h_{n,\ ch}^{(\al)}(\ep^2) \ =\ \frac{1}{\ep}\ h_{2n}^{(\al+1/2)}(\ep) \ ,
\label{hrel}
\ee
where the right hand side has to be taken at $2N$ as for the
polynomials themselves.
Consequently the same analysis for the "chiral" Cauchy transforms
follows from the same differential equation (\ref{chidiff}).

We note that while the Cauchy transform eq. (\ref{chCauchy}) defines a single
function with a cut along the positive real line $\mathbb{R_+}$,
the second transform in eq.  (\ref{chh}) defines again two functions in the
upper and lower half plane as in eq. (\ref{Cauchy}). It is in this latter
picture on  $\mathbb{R}$ that the results of the UE can be translated to the
chUE. This is why in the following section we will always give two
different solutions for the chiral Cauchy transforms as well.

\sect{Universality}\label{univ}

In this section we demonstrate the universality of
the asymptotic polynomials and
their Cauchy transform in the microscopic limit, which then
translates to all correlation functions
of characteristic polynomials through eq. (\ref{KN}).
The limit is called microscopic since it includes a certain rescaling of the
eigenvalues with an appropriate power of the parameter $N$,
according to the spectral region we investigate.
We will first study the microscopic limit at the origin (frequently
called the "hard edge"). This case is singled out due to the presence of
logarithms in the potential.
The correlations in the bulk of the spectrum then follow easily.
Furthermore we also study the edge of
the spectrum defined by the endpoints of the support of the macroscopic or
smoothed spectral density at large $N$ ("soft edge").
Here we will restrict ourselves to the cases
where the spectrum support is a simply connected
interval, the so-called one-cut phase. We will also stay away
from multicritical points, where the macroscopic density develops extra
zeros at the boundary of spectra or inside the support.
This amounts to requiring the limiting functions for $A_N(\la)$ to be
nonvanishing on the support as well (see eq. (\ref{Arho}) below).
While such multicritical
cases can be, in principle, analyzed in the framework of the same approach
\cite{ADMNII,KF97b,BE} the issue becomes much more subtle
due to subleading terms in the large-$N$ expansion.

Our first step is to derive the large-$N$ limit of the differential equations
(\ref{psidiff}) and (\ref{chidiff}). Our assumption of the one-cut support
is equivalent to the fact
that in the large-$N$ limit the recursion
coefficients $c_n$ in eq. (\ref{reccoeff}) approach a single valued function
\be
\lim_{n\to\infty}c_n=c(t=n/N)\ .
\label{cN}
\ee
While in the earlier physical matrix model literature
this assumption was introduced as an ansatz supported by numerical evidence,
it has been proved rigorously in the recent mathematical literature
\cite{Deift1}.
Eq. (\ref{cN}) implies the following asymptotic form
for the functions $A_N(\la)$ and $B_N(\la)$ in the differential
equation \cite{KFrev}:
\bea
\lim_{N\to\infty}\frac{1}{N} A_N(\la) &\equiv& A(\la) \ =\
\frac{a}{2\pi}\int_{-a}^a \!dt\ \frac{tV'(t)-\la V'(\la)}{t^2-\la^2}
\frac{1}{\sqrt{a^2-t^2}}
\label{Alim} \\
&=& \pi \rho(\la)
\frac{a}{\sqrt{a^2 - \la^2}} ~, \ \ \la\in\ (-a,a)\ .
\label{Arho}
\eea
Here $a\!=\! \lim_{N\to\infty}2c_N$ is the endpoint of the spectral support.
For convenience we have also presented
the relation to the macroscopic density of eigenvalues denoted by $\rho(\la)$.
Using the recursion relation eq. (\ref{rec}) it can be easily seen that
$A(\la)$ remains a polynomial of degree $d-2$, but the coefficients
get smoothened by taking the large-$N$ limit.
This corresponds to
the one cut phase through eq. (\ref{Arho}), with the nonpolynomial part of
$\rho(\la)\sim \sqrt{a^2-\la^2}$ being a simple square root.
The regular part $B_n^{\mbox{\scriptsize reg}}(\la)$
of the function $B_N(\la)$ also remains in the same limit a polynomial of
order $d-3$, with the limiting form \cite{KF}:
\bea
\lim_{N\to\infty}\frac{1}{N} B_N(\la) &\equiv&
B{\mbox{\scriptsize reg}}(\la) \ +\ (1-(-1)^N)
\frac{\al}{N\la} \nn\\
&=& \frac{1}{2\pi}\int_{-a}^a \!dt\ \frac{\la V'(t)-t V'(\la)}{t^2-\la^2}
\frac{1}{\sqrt{a^2-t^2}}\ +\ (1-(-1)^N)
\frac{\al}{N\la}
\label{Blim}
\eea
Using these expressions we obtain the following result for the
combination entering the differential equation eq. (\ref{ABV})
\begin{equation}
\lim_{N\to\infty}
\left(B_N(\la) + \frac{N}{2}V_{\al}'(\la)\right) \ =\
N\ \frac{\la}{2a}A(\la)-(-1)^N\
            \frac{\al}{\la} \ .
\label{ABlim}
\end{equation}
We have kept the last term as it will give rise to two different equations for
even and odd $N$.
Consequently, $G_N(\la)$ can be expressed asymptotically only in
terms of  $A(\la)$ and  therefore via
$\rho(\la)$ by virtue of Eq. (\ref{Arho}).
We have:
\begin{eqnarray}
\lim_{N\to\infty}F_N(\la) &=& \frac{A'(\la)}{A(\la)} \label{Flim}\\
\lim_{N\to\infty}G_N(\la) &=& N^2A(\la)^2\left(1-\frac{\la^2}{a^2}\right)+
                              \frac{(-1)^N\al-\al^2}{\la^2}
                +(-1)^N\ \frac{\al}{\la}\frac{A'(\la)}{A(\la)} \ .
\label{Glim}
\end{eqnarray}
Although it seems that the first term in $G_N(\la)$
is always dominating the
second term may contribute in some regimes as well,
depending on the chosen rescaling of
the eigenvalues which we are about to investigate in the next subsection.

\subsection{The hard edge}

We start by defining the microscopic rescaling in the large-$N$ limit
close to the hard-edge
origin of the spectrum to be
\be
N\la \ \equiv \ \xi\ .
\label{hardlim}
\ee
This amounts to magnifying the correlations at the point
$\la\approx 0$ as $\xi$ is kept fixed while $N\to\infty$.
Inserting this rescaling into the differential equations
(\ref{psidiff}) and (\ref{chidiff})
and using the asymptotic expressions for the coefficients,
eqs. (\ref{Flim}) and (\ref{Glim}) it is easy to see that the two terms
with the logarithmic derivative of $A(\la)$ are both of subleading order
${\cal O}(1/N)$. We thus obtain
\be
\varphi_N''(\xi) + \left( \pi^2\rho(0)^2+\frac{(-1)^N\al-\al^2}{\xi^2}\right)
\varphi_N(\xi)\ =\ 0 \ ,\ \ \varphi_N\in \{\psi_N,\ \chi_N\}\ .
\label{diffhard}
\ee
This expression is clearly universal as the dependence on all the
specific parameters $\{g_{2k}\}$ in the
polynomial potential in eq. (\ref{weightUE}) has been translated
into a single parameter - the macroscopic spectral density at
the origin $\rho(0)$. Due to the oscillating sign
we have to distinguish between even and odd values of
$N$, corresponding to even and odd polynomials and their Cauchy transforms.
The second order differential equation has two independent solutions
and it is the requirement of regularity of the solution at the origin
that dictates which one to be chosen.
The wave functions associated with the polynomials
$\psi_N(\xi)$ are regular at the
origin, fixing the corresponding solution to be Bessel functions of the first kind
$J_\nu(\xi)$. On the other hand the wave functions associated with the
Cauchy transforms are singular at the origin. As we will
see below this makes us to choose solutions to be
Bessel functions of the third kind also known as the Hankel functions,
$H^{(1)}_\nu(\zeta)$ or $H^{(2)}_\nu(\zeta)$, depending on the sign of
the imaginary part of the complex argument $N\ep\equiv\zeta$.
We therefore obtain
\bea
\psi_{2n}(\xi) &\sim&
(\pi\rho(0)\xi)^{\frac12}J_{\al-\frac12}(\pi\rho(0)\xi) \ ,\nn\\
\psi_{2n+1}(\xi) &\sim&
(\pi\rho(0)\xi)^{\frac12}J_{\al+\frac12}(\pi\rho(0)\xi) \ ,
\label{psiasymp}\\
&&\ \nn\\
\chi_{2n}(\zeta) &\sim&\left\{
\begin{array}{rl}
\ \ (\pi\rho(0)\zeta)^{\frac12}H^{(1)}_{\al-\frac12}(\pi\rho(0)\zeta) &\ , \ \
\im(\zeta)>0 \\
-(\pi\rho(0)\zeta)^{\frac12}H^{(2)}_{\al-\frac12}(\pi\rho(0)\zeta) &\ , \ \
\im(\zeta)<0 \\
\end{array}
\right. \ ,\nn\\
\chi_{2n+1}(\zeta) &\sim&\left\{
\begin{array}{rl}
\ \ (\pi\rho(0)\zeta)^{\frac12}H^{(1)}_{\al+\frac12}(\pi\rho(0)\zeta) &\ , \ \
\im(\zeta)>0  \\
-(\pi\rho(0)\zeta)^{\frac12}H^{(2)}_{\al+\frac12}(\pi\rho(0)\zeta) &\ , \ \
\im(\zeta)<0  \\
\end{array}\right. \ .
\label{chiasymp}
\eea

In order to extract from these wave-function asymptotics the
expressions for the bare polynomials and their Cauchy transforms
we still have to remove the corresponding weight factor from eqs.
(\ref{psi}) and (\ref{chi}) accordingly.
The microscopic limit of the weight function eq. (\ref{weightUE}) reads:
\be
\lim_{N\to\infty}N^{2\al}w(\la)\ =\ \xi^{2\al} \ .
\label{microw}
\ee
Multiplying or dividing the respective wavefunction
by the square root of the weight function,
we arrive at the final universal expressions to be inserted
into the determinant of eq. (\ref{KN}):
\bea
p_{2n}(\xi) &\sim&
\xi^{-\al+\frac12}J_{\al-\frac12}(\pi\rho(0)\xi)\ ,\nn\\
p_{2n+1}(\xi) &\sim&
\xi^{-\al+\frac12}J_{\al+\frac12}(\pi\rho(0)\xi)\ ,
\label{OPasymp}\\
&&\ \nn\\
h_{2n}(\zeta) &\sim& \left\{
\begin{array}{rl}
\ \ \zeta^{\al+\frac12}H^{(1)}_{\al-\frac12}(\pi\rho(0)\zeta) & \ , \ \
\im(\zeta)>0\\
-\zeta^{\al+\frac12}H^{(2)}_{\al-\frac12}(\pi\rho(0)\zeta)
& \ , \ \ \im(\zeta)<0\\
\end{array}
\right. \ ,\nn\\
h_{2n+1}(\zeta) &\sim& \left\{
\begin{array}{rl}
\ \ \zeta^{\al+\frac12}H^{(1)}_{\al+\frac12}(\pi\rho(0)\zeta) & \ , \ \
\im(\zeta)>0\\
-\zeta^{\al+\frac12}H^{(2)}_{\al+\frac12}(\pi\rho(0)\zeta)
& \ , \ \ \im(\zeta)<0\\
\end{array}
\right. \ .
\label{hasymp}
\eea
Let us now justify our choice of the type of Bessel functions
featuring in eq. (\ref{chiasymp}).
Due to the singularity at the origin we have to choose
between Bessel functions of second kind (also known as Weber
functions), $Y_\nu(\zeta)$,
and the Hankel functions. We give two independent arguments in favor of
the latter choice.
The first reason for selecting the Hankel functions is that it allows
one to match the known "bulk" asymptotic result \cite{FSuniv} after
taking the asymptotic limit $\zeta\to\infty$.
While the Weber functions would asymptotically produce sine
and cosine terms (as the Bessel functions of first kind do), the
Cauchy transforms in the bulk were shown to
approach $\exp[\pm i\zeta]$ \cite{FSuniv}.
This asymptotic is correctly reproduced by the Hankel functions.
The second argument is related with the necessity to distinguish between
the two types of the Hankel functions $H^{(1)}_\nu(\zeta)$ and
$H^{(2)}_\nu(\zeta)$ according to the sign of $\im(\zeta)$.

In fact, for $\al=0$ and Gaussian potential $w(\la)=\exp[-N\la^2]$
one can explicitly write down the integral representations
for the polynomials and Cauchy transforms at finite-$N$, as discussed
in \cite{FA}.
Switching back to the monic normalization the polynomials
are the Hermite polynomials $\pi_n(\la)=H_n(\la\sqrt{N})/(2\sqrt{N})^n$.
It follows that
\bea
\pi_{2m}(\la)   &=& \mbox{e}^{N\la^2}(-1)^m 2N\sqrt{\la}\int_0^\infty ds\
  \mbox{e}^{-Ns^2} s^{2m+\frac12} J_{-\frac12}(2Ns\la) \nn\\
\pi_{2m+1}(\la) &=&  \mbox{e}^{N\la^2}(-1)^m 2N\sqrt{\la}\int_0^\infty ds\
\mbox{e}^{-Ns^2} s^{2m+\frac32} J_{+\frac12}(2Ns\la) \ ,
\label{OPrepherm}
\eea
where we have used the standard integral representation of the Hermite
polynomials as well as reexpressed cosine and sine by the half-integer
Bessel functions
$J_{\pm\frac12}$. As a consistency check one can verify
that eq. (\ref{OPasymp}) at $\al=0$
follows from a saddle point approximation in the microscopic large-$N$ limit,
remembering that $\pi\rho(0)=\sqrt{2}$ for the Gaussian potential.
Using eq. (\ref{OPrepherm}) we obtain a finite-$N$ integral representation of
the Cauchy transforms as follows. We can rewrite
\be
\frac{1}{x-\ep}\ =\ \mbox{sgn}[\im(\ep)]\,i \int_0^\infty dt\
\exp[-\mbox{sgn}[\im(\ep)]it(x-\ep)]
\ ,\ \ \im(\ep)\neq 0\ .
\label{auxint}
\ee
Inserting eqs. (\ref{OPrepherm}) and (\ref{auxint}) into the definition
(\ref{Cauchy}) we can first perform the integral over $\la$ and then
the one over $t$, arriving at
\bea
\vartheta_{2m}(\ep) &=& (-1)^m N\sqrt{\ep}
\int_0^\infty ds\ \mbox{e}^{-Ns^2}s^{2m+\frac{1}{2}}
\left\{
\begin{array}{rl}
\ \ H^{(1)}_{-\frac12}(2Ns\ep)& \ , \ \ \im(\ep)>0\\
-H^{(2)}_{-\frac12}(2Ns\ep)& \ , \ \ \im(\ep)<0\\
\end{array}\ ,
\right. \nn\\
\vartheta_{2m+1}(\ep) &=& (-1)^m N\sqrt{\ep}
\int_0^\infty ds\ \mbox{e}^{-Ns^2}s^{2m+\frac{3}{2}}
\left\{
\begin{array}{rl}
\ \ H^{(1)}_{+\frac12}(2Ns\ep)& \ , \ \ \im(\ep)>0\\
-H^{(2)}_{+\frac12}(2Ns\ep)& \ , \ \ \im(\ep)<0\\
\end{array}
\right. \ .
\label{Cauchyrep}
\eea
In taking the microscopic large-$N$ limit we again recover
eq. (\ref{hasymp}) at $\al=0$ from a saddle point approximation.
This fact confirms the correct choice of the solution for the
Cauchy transform and thus for the wavefunctions eq. (\ref{chiasymp})
for a Gaussian potential with $\al=0$. Since the parameters
$\al$ as well as the higher order coupling constants
$g_{2k}$ in the potential eq. (\ref{Valpha}) are real and can be switched
on smoothly the choice of the type of Hankel function remains fixed
for a general potential for general $\al$ as well.

\indent

Now we can use the relations between the polynomials and
Cauchy transforms of the
chiral and non-chiral ensemble, eqs. (\ref{OPrel}) and (\ref{hrel})
respectively, to read off immediately the universal asymptotic formulas
for the {chiral Unitary Ensemble} from (\ref{OPasymp}) and (\ref{hasymp}):
\bea
p_{n,\ ch}^{(\al)}(\xi^2) &\sim&
\xi^{-\al}J_{\al}(\pi\rho(0)\xi)\ ,
\label{chOPasymp}\\
&&\ \nn\\
h_{n,\ ch}^{(\al)}(\zeta^2) &\sim& \left\{
\begin{array}{rl}
\ \ \zeta^{\al}H^{(1)}_{\al}(\pi\rho(0)\zeta) & \ , \ \
\im(\zeta)>0\\
-\zeta^{\al}H^{(2)}_{\al}(\pi\rho(0)\zeta)
& \ , \ \ \im(\zeta)<0\\
\end{array}
\right. \ .
\label{chhasymp}
\eea
We see that for integer values of $\al$
the indices of the Bessel functions
are also integer, compared to half
integer values for the non-chiral case above.
To fix independently the correct
Bessel solution corresponding to the sign of
$\im(\xi)$ we can repeat the argument as in eqs. (\ref{OPrepherm}) and
(\ref{Cauchyrep}), but this time for general $\alpha\neq 0$.
The corresponding finite-$N$ expression have already been given in
\cite{FA}, following the integral representation
of generalized Laguerre polynomials, and we only repeat them here:
\bea
\pi_{n,\,ch}(x^2)&=& \mbox{e}^{Nx^2} (-1)^n Nx^{-\al}
\int_0^{\infty}dt\, \mbox{e}^{-Nt}t^{n+\al/2}J_{\al}
\left(2Nx\sqrt{t}\right)\ ,
\nn\\
\vartheta_{n,\, ch}(\epsilon^2)    &=&\frac{(-1)^n}{2}
N\epsilon^{\al}
\int_0^{\infty}dt\, \mbox{e}^{-Nt}t^{n+\al/2}
 \left\{
\begin{array}{rl}
\ \ H^{(1)}_{\al}\left(2N\epsilon \sqrt{t}\right)& \ , \ \ \im(\ep)>0\\
-H^{(2)}_{\al}\left(2N\epsilon \sqrt{t}\right)& \ , \ \ \im(\ep)<0\\
\end{array}
\right. \ .
\label{chintrep}
\eea
The  saddle point approximation reconfirms eq. (\ref{chhasymp}) for the
Gaussian potential $V_\al^{ch}(\la)=\la-\frac{\al}{N}\ln(\la)$
and, by invoking the continuity argument for a general potential as well.

\subsection{The bulk of the spectrum}

In this section we briefly describe the microscopic limit
in the bulk of the spectrum, which
amounts to setting $\al=0$ in the differential equation
(\ref{diffhard}). For the non-chiral case the corresponding formulae
match with those obtained in \cite{FSuniv}.

It is evident that when logarithmic terms are absent the
origin of the spectrum is no longer singled out. Under these conditions
correlation functions when properly scaled
should be the same at every point inside the support of the spectrum,
due to translational invariance.
More precisely, let us pick a fixed value $\la_*\in (-a,a)$, that means
neither at the origin nor at the edge of the support.
The microscopic rescaling around the chosen point is
then defined as
\be
N(\la-\la_*)\ \equiv\ \xi\ .
\label{bulklim}
\ee
Inserting this expression into that for the asymptotic coefficients,
eqs. (\ref{Flim}) and
(\ref{Glim}), we can easily see that only the first term in
$G_N(\la)$ contributes yielding
\be
\varphi_N''(\xi) +  \pi^2\rho(\la_*)^2
\varphi_N(\xi)\ =\ 0 \ ,\ \ \varphi_N\in \{\psi_N,\ \chi_N\}\ .
\label{diffbulk}
\ee
where we used eq. (\ref{Arho}) squared.
This is just eq. (\ref{diffhard}) with
$\al=0$, replacing $\rho(0)$ by the universal parameter $\rho(\la_*)$. From
eqs. (\ref{psiasymp}) and
(\ref{chiasymp}) we see that the solutions are Bessel functions of index
$\pm\frac{1}{2}$, which are just sine and cosine.
The differential equation no longer
distinguishes between even and odd $N$, but the parity fixes the solution
uniquely. The solutions for the asymptotic Cauchy
transform will no longer be singular as we expand around the point
$\la=\la_*$. However, we can still use the results of the previous subsection
for $\al=0$ in order to fix the solution uniquely
depending on the sign $\im(\ep)$ in the Cauchy transform.
Omitting the constant proportionality factor
$w(\la_*)^{\pm\frac12}$ when switching from
wavefunctions to polynomials and their Cauchy transforms
yields the explicit expressions for the latter:
\bea
p_{2n}(\xi) &\sim& \cos(\pi\rho(\la_*)\xi)\ ,\nn\\
p_{2n+1}(\xi) &\sim&
\sin(\pi\rho(\la_*)\xi)\ ,
\label{OPbulkasymp}\\
&&\ \nn\\
h_{2n}(\zeta) &\sim& \left\{
\begin{array}{rl}
\exp(+i\pi\rho(\la_*)\zeta)
& \ , \ \ \im(\zeta)>0\\
-\exp(-i\pi\rho(\la_*)\zeta)
& \ , \ \ \im(\zeta)<0\\
\end{array}
\right. \ ,\nn\\
h_{2n+1}(\zeta) &\sim& \left\{
\begin{array}{rl}
-i\ \exp(+i\pi\rho(\la_*)\zeta)
& \ , \ \ \im(\zeta)>0\\
-i\ \exp(-i\pi\rho(\la_*)\zeta)
& \ , \ \ \im(\zeta)<0\\
\end{array}
\right. \ .
\label{hbulkasymp}
\eea

Due to the fact that no $\al$-dependence is present in eq. (\ref{diffbulk})
the polynomials and Cauchy transform in the chiral ensemble are identical to
the even half of the non-chiral ones.  The eqs. (\ref{OPrel}) and
(\ref{hrel}) then give:
\bea
p_{n,\ ch}^{(\al)}(\xi^2) &\sim&
 \cos(\pi\rho(\la_*)\xi)\ ,
\label{chOPbulkasymp}\\
&&\ \nn\\
h_{n,\ ch}^{(\al)}(\zeta^2) &\sim& \left\{
\begin{array}{rl}
\zeta^{-1}\exp(+i\pi\rho(\la_*)\zeta)
& \ , \ \ \im(\zeta)>0\\
-\zeta^{-1}\exp(-i\pi\rho(\la_*)\zeta)
& \ , \ \ \im(\zeta)<0\\
\end{array}
\right. \ .
\label{chhbulkasymp}
\eea

\subsection{The soft edge}

Finally we turn our attention to the correlation functions
in the vicinity of the edges $\la=\pm a$ of
the support of the spectrum, the so-called soft edges \cite{soft,BHsoft}.
Restricting ourselves to $\la=+a$ for symmetry reasons we define
the microscopic scaling limit at the soft edge as
\be
N^{\frac23}(\la-a)\ \equiv\ \xi \ .
\label{edgelim}
\ee
In this way we magnify the region around the endpoint of the support.
For negative (positive) $\xi$ we are inside (outside) the support.
In the limit eq. (\ref{edgelim}) only the first term in eq. (\ref{Glim})
contributes and the different power in $N$ is due to the factor $(a^2-\la^2)$
that multiplies $A(\la)^2$. In the large-$N$ limit the differential
equation acquires the form:
\be
\varphi_N''(\xi) -  \xi \frac{2A(a)^2}{a}
\varphi_N(\xi)\ =\ 0 \ ,\ \ \varphi_N\in \{\psi_N,\ \chi_N\}\ .
\label{diffedge}
\ee
It is again universal, extending the results of \cite{KF97b} to the
Cauchy transforms.
After changing variables the solutions is obtained  in terms of
the Airy functions. Again omitting the prefactors $w(a)^{\pm\frac12}$ when
changing
from wavefunctions to the original polynomials and Cauchy transforms
we obtain:
\bea
p_{N}(\xi) &\sim& \mbox{Ai}\left(\left(2a^{-1}A(a)^2\right)^{\frac13}\xi\right)
\ ,\label{OPedgeasymp}\\
&&\ \nn\\
h_{N}(\zeta) &\sim&
\left\{
\begin{array}{rl}
-\,\mbox{e}^{+\frac{2\pi i}{3}}
\mbox{Ai}\left( (2a^{-1}A(a)^2)^{\frac13}\zeta\,
\mbox{e}^{+\frac{2\pi i}{3}}\right) & \ , \ \ \im(\zeta)>0\\
\ \ \,\mbox{e}^{-\frac{2\pi i}{3}}
\mbox{Ai}\left( (2a^{-1}A(a)^2)^{\frac13}\zeta\,\mbox{e}^{-\frac{2\pi i}{3}}
\right) & \ , \ \ \im(\zeta)<0\\
\end{array}
\right.\ .
\label{hedgeasymp}
\eea
Here we have chosen for the Cauchy transforms one of the two solutions
of the Airy equation, that one which is linearly independent
of $\mbox{Ai}(x)$ \cite{abramo}, in coordination with
the sign of $\im(\zeta)$:
\be
\mbox{Ai}\left( z\,\mbox{e}^{\pm\frac{2\pi i}{3}}\right) \ =\
\frac12  \,\mbox{e}^{\pm\frac{\pi i}{3}}
\left( \mbox{Ai}(z) \ \mp i\, \mbox{Bi}(z)\right)
\label{A's}
\ee

The functions Bi$(z)$ is a new feature emerging when
negative moments of the characteristic polynomials
are considered. For positive moments only \cite{BHsoft}, as well as for the
spectral densities and related objects \cite{Mehta,soft,FW}
only functions Ai$(z)$ feature in the final results.

The discontinuity between the two solutions is again a simple Airy function,
due to the functional relation \cite{abramo}:
\be
\mbox{Ai}(z)\ =\ -\,\mbox{e}^{+\frac{2\pi i}{3}}
\mbox{Ai}\left( z\,\mbox{e}^{+\frac{2\pi i}{3}}\right) \ -\
\mbox{e}^{-\frac{2\pi i}{3}}
\mbox{Ai}\left( z\,\mbox{e}^{-\frac{2\pi i}{3}}\right)
\ee
For real negative argument
the Airy functions $\mbox{Ai}$ and $\mbox{Bi}$
are oscillating while for positive
argument they monotonously decrease/increase, respectively.
For large positive argument we conclude that
\be
\lim_{z\to+\infty}\mp\,\mbox{e}^{\pm\frac{2\pi i}{3}}
\mbox{Ai}\left( z\,\mbox{e}^{\pm\frac{2\pi i}{3}}\right)
\ =\ -i \,\mbox{Bi}(z) \ ,
\label{equalh}
\ee
and thus see that both solutions eq. (\ref{hedgeasymp}) agree.
This fixes our choice of the boundary condition as for large
positive arguments outside the support the Cauchy transforms should no
longer experience a discontinuity across the real line.

The solution for even and odd $N$ is eqs.  (\ref{OPedgeasymp}) and
(\ref{hedgeasymp}) are the same. This immediately
implies that the results for the UE and the
chUE are identical, as must be the case far from the origin of the spectrum.
Therefore we have to compute also the
difference between $\varphi_N$ and $\varphi_{N-1}$ as
they both appear in expressions relating the correlation functions containing
ratios  of characteristic polynomials to orthogonal polynomials and
their Cauchy transform\footnote{The same
remark in fact applies already to the chUE in the
previous subsections.}.
For that aim we use explicitly the differentiation eqs. (\ref{Pprime}) and
(\ref{hprime}) for the corresponding wave functions together with the
limiting relation eq. (\ref{Blim}).
Following \cite{KFrev} we finally obtain:
\be
\varphi_{N-1}(\xi)\ =\ \varphi_N(\xi)
\ + \ \frac{1}{N^{\frac13}A(a)}\varphi_N'(\xi)
\ ,\ \ \varphi_N\in \{\psi_N,\ \chi_N\}\ .
\label{correction}
\ee

This completes our demonstration of universality for the
orthogonal polynomials and their Cauchy transforms for both the UE
and chUE, in all three different regions of the spectrum.

We close with the following comment.
Once the  universality of
expectation values of ratios of characteristic polynomials
eq. (\ref{defKN}) is demonstrated it
immediately translates to that of all other
correlation functions
that are generated by them by differentiations.
As a check we can consider the correlation function of the
 eigenvalue densities by using
eqs. (\ref{OPasymp}) - (\ref{hasymp})
eqs. (\ref{chOPasymp}) - (\ref{chhasymp}) respectively,
and taking necessary derivatives and discontinuities.
In this way one reproduces the known universal
microscopic spectral densities of the unitary and
chiral unitary ensemble, respectively, see e.g. \cite{ADMN}.
It covers the bulk limit by setting $\al=0$.
The difference is that in the chUE the resolvent
has a true singularity at the origin while in the UE the discontinuity
originates from patching two different solutions, each of which is regular
at this point. This is another way of seeing that for the UE the
``old'' replica limit using saddle point techniques is only able
to reproduce the asymptotic expansions \cite{DV}
while for the chiral case it yields the exact result \cite{ADDV}.
Similar checks for the universal microscopic eigenvalue density
at the soft edge can be performed using  eqs. (\ref{OPedgeasymp}) and
(\ref{hedgeasymp}), recovering known expression from \cite{KF97b}.


\sect{Discussion}\label{con}

For a broad class of measures we have shown
that the Cauchy transforms of orthogonal polynomials
display properties very similar to those of the polynomials themselves.
In particular, they obey the same three step recursion relation.
The wave functions associated with the Cauchy transforms
satisfy the same second order differential equation
as those associated with the polynomials.
This correspondence holds for finite-$N$
with polynomial potentials of arbitrary
but fixed even degree $d$, provided that $N>d-2$.
In the microscopic large-$N$ limit we have thus obtained the same
universal Bessel equation at the hard edge, and the Airy equation at
the soft edge of the spectrum.
For the Cauchy transform the choice of the correct solutions is dictated
by their singularity at the origin and the sign of the imaginary part
of the argument. This identification was also supported
by a comparison with exact integral representations
available for the Gaussian weight at finite-$N$.

Using the general results \cite{FS}
the universality of arbitrary ratios of characteristic
polynomials in RMT with unitary invariance thus follows.
Our consideration complements the rigorous
proof \cite{FSuniv} valid for the bulk of the spectrum and
relied on the Riemann-Hilbert approach.
The arguments presented here rely on the previous results in
\cite{KFrev,Deift1}. Although partly heuristic
by strict mathematical standards, the method
has the  clear advantage of being much more transparent
and able to treat both cases of the chUE and the UE on equal footing.

Since our differential equation for the wavefunctions is valid at finite-$N$
we could be tempted to take the macroscopic large-$N$ limit as well.
It gives the eigenvalue correlations at distances large compared to the mean
level spacing, where the local fluctuations are smoothed.
While the spectral density itself is non-universal, all higher order
connected correlation functions are known to be universal
\cite{AJM}. It has been shown in \cite{KFrev} that the density and the
connected two point function can be also obtained from smoothing products
of orthogonal polynomials. However, for higher order correlations
the corrections to the orthogonal polynomials
become extremely difficult to evaluate and are unknown so far.
For that reason  we have not tried to consider smoothened
Cauchy transforms in order to recover the known
one and two-point function.
It is also less clear if the procedures of smoothening and
taking the discontinuity after differentiation can be interchanged
safely.

Finally it would be very interesting to extend the general results
\cite{FS} to matrix models with complex eigenvalues and to prove their
universality. First results in this direction have already been obtained
\cite{AVIII}. There,
arbitrary products of characteristic polynomials have been
expressed in a determinant formula, which allowed to demonstrate
 their universality at the regime of weak non-Hermiticity.

\indent

\noindent
\underline{Acknowledgments}:
We wish to thank Graziano Vernizzi and Boris Khoruzhenko for a
critical reading and comments on the manuscript.
This work was supported by Brunel University Vice Chancellor Grant
(YVF) and by the Heisenberg fellowship
of the Deutsche Forschungsgemeinschaft (GA).


\indent

\end{document}